# Observation of flat-bottom U-shaped energy gap in high-$T_c$ nickelate $(La,Pr)_3Ni_2O_7$ thin films

Zhen Liang[1,2,3,#], Tianheng Wei[1,2,#], Wei Ren[1,2,#], Haoran Ji[1,2], Zheyuan Xie[1,2], Yanzhao Liu[4], Ziqiang Wang[5] & Jian Wang[1,2,3,6,*]

[1]International Center for Quantum Materials, School of Physics, Peking University, Beijing 100871, China

[2]Beijing Key Laboratory of Quantum Devices, Peking University, Beijing 100871, China

[3]Hefei National Laboratory, Hefei 230088, China

[4]Quantum Science Center of Guangdong-Hong Kong-Macao Greater Bay Area, Shenzhen, 518045, China

[5]Department of Physics, Boston College, Chestnut Hill, MA 0246, USA

[6]Collaborative Innovation Center of Quantum Matter, Beijing 100871, China

[#]These authors contributed equally.

[*]Corresponding to: jianwangphysics@pku.edu.cn (J.W.).

**The discovery of high transition temperature (high-$T_c$) superconductivity in Ruddlesden-Popper (R-P) bilayer nickelates under high pressure[1–3] has stimulated extensive work to understand the underlying mechanism and search for superconductors with higher $T_c$[4–22]. The recent realization of superconductivity in R-P bilayer nickelate thin films with onset $T_c$ above 40 K at ambient-pressure enables the use of a wide array of powerful experimental tools to investigate the unconventional high-$T_c$ superconductivity in bilayer nickelates[23–26]. Here, using ultra-low temperature scanning tunneling microscopy/spectroscopy (STM/S) and electrical transport study, we report the first successful observation of an energy-symmetric, flat-bottom U-shaped gap with zero residual density of states around the Fermi level in the high-$T_c$ nickelate $(La,Pr)_3Ni_2O_7$ thin film grown on $SrLaAlO_4$ substrate. Before and after STM/S studies, transport measurements on the same sample reveal consistent superconducting behaviors showing zero resistance, with an onset $T_c$ above 40 K and zero resistance $T_c$ above 20 K. The tunneling spectra exhibit highly unconventional temperature evolution, characterized by a rapid filling of the U-shaped energy gap to a V-shaped gap as the temperature increases. Furthermore, the U-shaped energy gap is reduced under a *c*-axis magnetic field of 14 T. The energy-symmetric U-shaped gap, taken together with its dependence on magnetic field and temperature, is consistent with the behavior of a superconducting gap, suggesting a nodeless gap function at ultra-low temperatures. Our findings shed new lights on the nature of high-$T_c$ superconductivity and provide an encouraging and thought-provoking hint for a local superconductivity with $T_c$ above liquid nitrogen boiling temperature in nickelate superconductors at ambient or zero pressure.**

Since the discovery of high transition temperature (high-$T_c$) superconductivity with an onset $T_c$ near 80 K in the Ruddlesden-Popper (R-P) bilayer nickelate $La_3Ni_2O_7$ under high pressure[1,2], the R-P nickelates have attracted significant research interest as a novel family of unconventional high-$T_c$ superconductors[4–22]. Nevertheless, the demanding high-pressure condition severely limits the feasibility of conventional experimental setups, thereby restricting extensive experimental investigations and potential applications. Recently, ambient-pressure superconductivity with onset $T_c$ above 40 K (the McMillan limit)[27,28] has been realized in R-P bilayer nickelate thin films through compressive strain from the substrates[23–26]. This provides unprecedented opportunities for comprehensive and in-depth studies of superconducting (SC) nickelates, unlocking the well-established experimental toolbox for studying superconductivity.

As a powerful experimental technique, scanning tunneling microscopy/spectroscopy (STM/S) has been widely used to investigate the high-$T_c$ cuprates[29], iron-based superconductors[30,31], as well as other unconventional superconductors[32,33]. The SC gap can be clearly revealed by the single-particle tunneling spectra, which probe the density of states of quasiparticle excitations in superconductors[29,34]. In this work, using ultra-low temperature STM/S, we report the discovery of an energy-symmetric, flat-bottom U-shaped gap with coherence peak structures at the gap edges in the bilayer nickelate $(La,Pr)_3Ni_2O_7$ thin film grown on $SrLaAlO_4$ substrate. The gap size $\Delta$, determined by one half of the distance between coherence peaks, is larger than 40 meV. Both the gap size and the coherence peak structures are reduced by increasing temperature or applying a magnetic field, consistent with the evolution of a SC gap. The spectral weight is completely suppressed over a finite range around the Fermi level at ultra-low temperatures, suggesting the gap is nodeless. Moreover, the exceptionally large U-shaped gap is rapidly filled and evolves into a V-shaped gap when the temperature increases from 60 mK to 10 K, which is strikingly small compared to both the gap size $\Delta$ (41.6 meV) and the $T_c$ (onset $T_c$ above 40 K and zero resistance $T_c$ above 20 K) measured by transport experiments. Our experimental studies provide new insights for understanding the unconventional superconductivity in the bilayer nickelates, and indicate the potential of R-P bilayer nickelate $(La,Pr)_3Ni_2O_7$ thin films to achieve superconductivity above the liquid nitrogen boiling temperature at ambient pressure.

**Flat-bottom U-shaped energy gap with energy-symmetric coherence peak structures in the bilayer nickelate thin film**

The three-unit-cell thick $(La,Pr)_3Ni_2O_7$ thin film (Sample 1) is grown on the as-received (001)-oriented $SrLaAlO_4$ substrate by gigantic-oxidative atomic layer-by-layer epitaxy (GAE)[24,35]. As a member of the R-P bilayer nickelates, $(La,Pr)_3Ni_2O_7$ has the stacking structure of $(La,Pr)O\text{-}NiO_2\text{-}(La,Pr)O\text{-}NiO_2\text{-}(La,Pr)O$, which is schematically shown in Fig. 1a. The La:Pr ratio of the $(La,Pr)_3Ni_2O_7$ thin film is about 1.95:1.05. The resistance-temperature ($R$-$T$) curve obtained by transport measurements (Fig. 1b) shows superconductivity with an onset SC $T_c$ of 46.3 K and zero resistance $T_c$ of 21.7 K. The measurement configuration using standard four-probe method is shown in the inset of Fig. 1b. After the transport measurements, the nickelate thin film (Sample 1) was transferred to the STM chamber. The surface of the $(La,Pr)_3Ni_2O_7$ thin film shows roughness on the nanometer scale (see Supplementary Information), which may be induced by the exposure to air when transferring the $(La,Pr)_3Ni_2O_7$ film to the STM chamber. Similar rough surface of the $(La,Pr)_3Ni_2O_7$ thin film was observed in a recent report[36]. To obtain a fresh surface, we treat the $(La,Pr)_3Ni_2O_7$ thin film by pushing the tip towards the sample for several nanometers (~ 8 nm) and then quickly pulling back, which may refresh the sample surface. After the tip treatment, the surface in a local region of the $(La,Pr)_3Ni_2O_7$ thin film is refreshed. The inset of Fig. 1c shows the STM topography of the $(La,Pr)_3Ni_2O_7$ thin film after the tip treatment over a measured region (Region 1). Despite the STM measured topography shows some disorders and slight discontinuity, the superconducting gap structure could still be revealed by tunneling spectra[36–38]. To investigate the electronic structure and superconductivity, we carried out the STS measurements. Figure 1c displays a typical tunneling spectrum measured at the ultra-low temperature $T$ = 60 mK at the location marked by the red star in the inset of Fig. 1c. Two coherence peaks and a flat-bottom U-shaped gap symmetric in energy around the Fermi level can be clearly observed in the tunneling spectrum. The gap size ($\Delta$) is around 41.6 meV, determined by one half of the separation between the two coherence peaks. Since the superconductivity of the $(La,Pr)_3Ni_2O_7$ thin film is realized through the compressive strain induced from the $SrLaAlO_4$ substrate[24], the observed extremely large gap cannot be ascribed to the scenario that the treatment process might bring part of the film to the tip thereby inducing superconductivity on the tip. We also carried out STS measurements at other positions in Region 1, as shown in Extended Data Fig. 1. The tunneling spectra are inhomogeneous at different sites, possibly due to disorders[39] and potential electronic glassiness[40] in the film. Despite the inhomogeneity, the flat-bottom U-shaped energy gap is universal in this region. Furthermore, to verify that our experimental results are reproducible, we carried out STM/S measurements in another region (Region 2) of the $(La,Pr)_3Ni_2O_7$ thin film after tip treatment using the same method (Extended Data Fig. 2), where the tunneling spectra also exhibit a flat-bottom U-shaped energy gap of a similar size. After STM/S experiments, we performed transport measurements again on the same nickelate thin film (Sample 1) and the $R$-$T$ curve is shown in Fig. 1d. The nickelate thin film still shows high-$T_c$ superconductivity after STM/S experiments, with an onset $T_c$ of 48.4 K and zero resistance $T_c$ of 23.0 K.

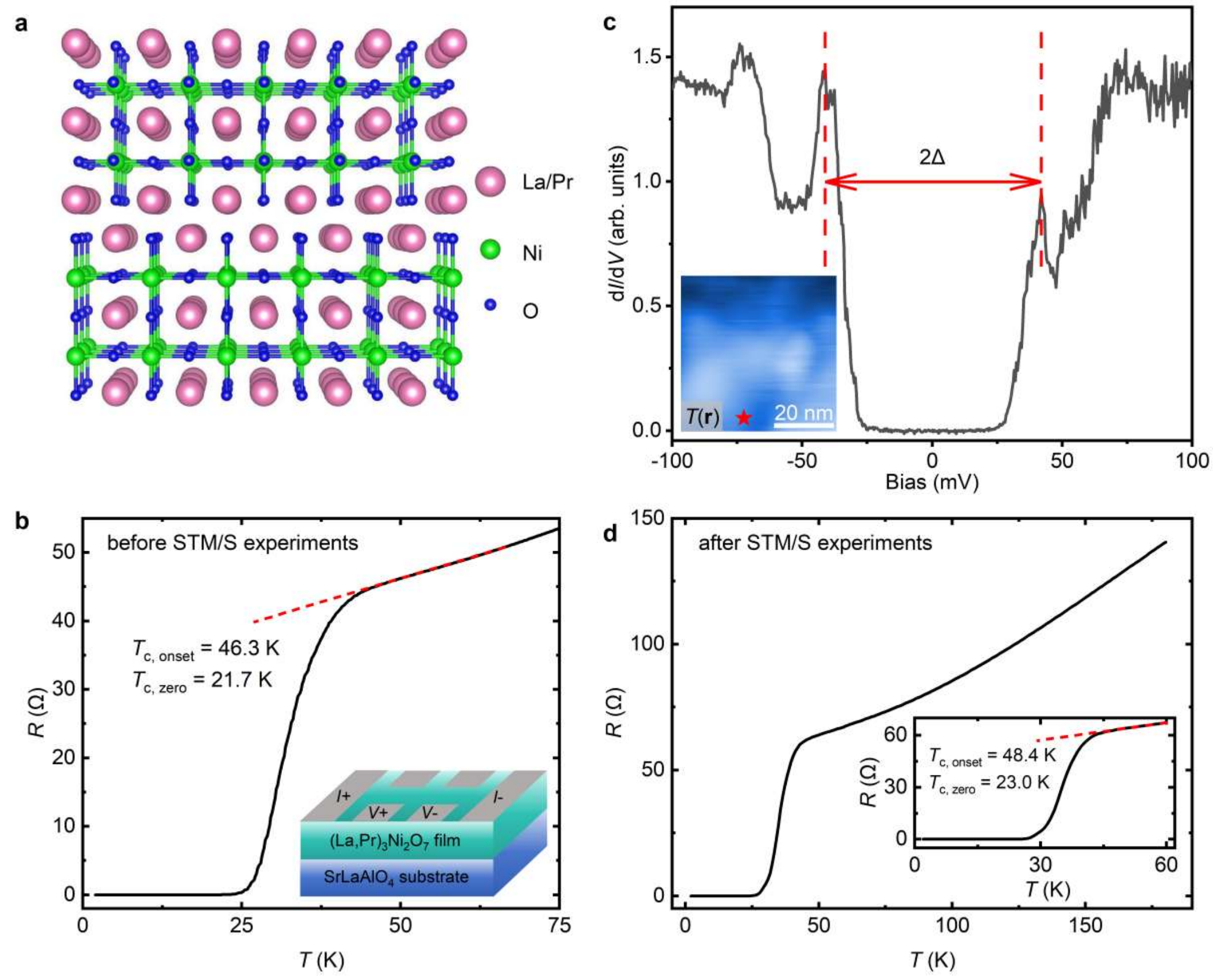

**Fig. 1 | SC nickelate (La,Pr)$_3$Ni$_2$O$_7$ thin film grown on SrLaAlO$_4$ substrate. a**, The schematic of the (La,Pr)$_3$Ni$_2$O$_7$ crystal structure. **b**, The *R*-*T* curve of the (La,Pr)$_3$Ni$_2$O$_7$ thin film measured before STM/S experiments. The onset SC $T_c$ is 46.3 K and the zero resistance $T_c$ is 21.7 K. The inset shows the measurement configuration. **c**, A typical tunneling spectrum measured at the position marked by the red star in the inset ($V_s$ = 100 mV, $I_s$ = 500 pA, $V_{mod}$ = 2 mV, $T$ = 60 mK). The gap size (Δ) is around 41.6 meV determined by one half of the separation in energy between the two coherence peaks. The inset shows an STM topographic image of the (La,Pr)$_3$Ni$_2$O$_7$ thin film in Region 1 (52.25 × 52.25 nm$^2$, $V_s$ = 8 V, $I_s$ = 100 pA) after the tip treatment. **d**, The *R*-*T* curve of the (La,Pr)$_3$Ni$_2$O$_7$ thin film measured after STM/S experiments. The inset is the zoom-in *R*-*T* curve, showing the onset SC $T_c$ of 48.4 K and the zero resistance $T_c$ of 23.0 K.

**Reduction of the U-shaped energy gap under the magnetic field**

To further investigate the origin of the U-shaped energy gap in the (La,Pr)$_3$Ni$_2$O$_7$ thin film, we carried out STS measurements along the red arrow shown in Fig. 2a, which is the STM topography measured in Region 1 (different area with the inset of Fig. 1c) at ultra-low temperature $T$ = 50 mK. The tunneling spectra (Fig. 2b) acquired at $T$ = 50 mK at different positions exhibit some differences due to the inhomogeneity of the film. Despite the inhomogeneity, all the tunneling spectra exhibit the U-shaped energy gap of the similar size. Then, we applied a magnetic field $B$ = 14 T perpendicular to the sample surface and carried out STS measurements along the same line-cut as shown in Fig. 2c. The gap structures survive under $B$ = 14 T. Since the tunneling spectra in the $(La,Pr)_3Ni_2O_7$ thin film shows real space inhomogeneity, in Fig. 2d, we compare the spatially averaged tunneling spectra at $B$ = 0 T (Fig. 2b) and $B$ = 14 T (Fig. 2c) to show the response to the magnetic field. The averaged energy gap measured under $B$ = 14 T is clearly reduced comparing to that measured in zero magnetic field, which is consistent with the evolution of a SC gap with extremely large critical field such as a high-$T_c$ superconductor. Note that the zero-resistance state of compressively strained bilayer nickelate thin films persists under 14 T at 2 K revealed by transport measurements[40]. In another region (Region 3) of the film, the suppression of coherence peaks is observed under the magnetic field (Extended Data Fig. 4), which is also consistent with the reduction of the superconductivity by applying magnetic field.

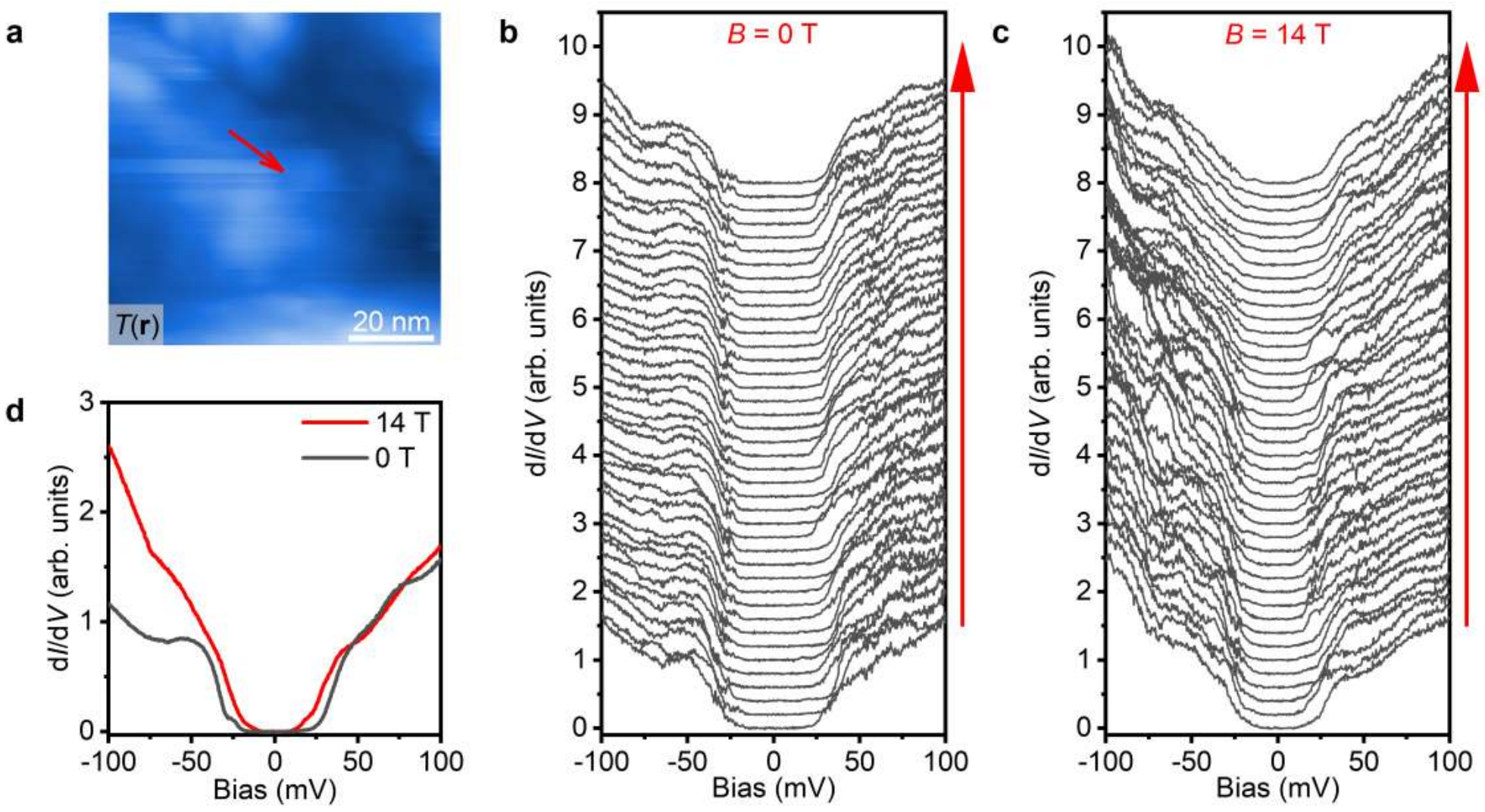

**Fig. 2 | U-shaped energy gap structure and reduction of the energy gap under magnetic field $B$ = 14 T perpendicular to sample surface. a**, A topographic image of the $(La,Pr)_3Ni_2O_7$ thin film in Region 1 (77.71 × 77.71 nm$^2$, $V_s$ = 8 V, $I_s$ = 100 pA). **b**, The tunneling conductance spectra measured along the red arrow shown in **a** without magnetic field ($V_s$ = 100 mV, $I_s$ = 500 pA, $V_{mod}$ = 2 mV, $T$ = 50 mK). The spectra are vertically shifted for clarity. **c**, The tunneling spectra measured along the same arrow in **a** under a magnetic field $B$ = 14 T perpendicular to the surface ($V_s$ = 100 mV, $I_s$ = 500 pA, $V_{mod}$ = 2mV, $T$ = 50 mK). The spectra are vertically shifted for clarity. **d**, The spatially average of the tunneling spectra in **b** ($B$ = 0 T) shown by the black curve and **c** ($B$ = 14 T) shown by the red curve. The energy gap is reduced by the magnetic field.

**Unconventional temperature evolution of the tunneling spectra**

In addition to the reduction of the U-shaped energy gap by magnetic fields, we also studied the temperature evolution of the energy gap. Figure 3a shows the tunneling spectra measured at different temperatures at the same position marked by the red star in the topography shown in the inset of Fig. 1c. As the temperature increases, the energy gap decreases. Remarkably, the complex evolution of the tunneling spectra with temperature is highly unconventional. Increasing the temperature from 60 mK to 7 K, the tunneling spectra exhibit rapid filling of the U-shaped gap and suppression of the coherence peaks (Fig. 3b). Starting from 10 K, the U-shaped gap elevates its bottom above zero and evolves into a V-shaped gap structure (Fig. 3c) with nonzero density of states at the Fermi level. The spectral weight transfer with increasing temperature is not limited around the energy gap, but rather occurs in a wider energy range (at least within ±100 meV), indicating the possible unconventional superconductivity of the bilayer nickelate thin film[41]. The V-shaped gap structure evolves only gradually with further increasing temperatures and can still be observed at $T$ = 45 K. The observed temperature evolution of the gap structure is much different from that of the high-$T_c$ cuprate superconductors, where the coherence peaks of the V-shaped SC gap are gradually weakened as the temperature increases toward $T_c$, and above $T_c$, the pseudogap structure can be observed until a much higher temperature[42].

To illustrate that the highly unconventional temperature-dependence and suppression of the gap structure do not simply result from the thermal broadening effect, we convolute the spectrum measured at 60 mK by the thermal broadening functions at higher temperatures (Methods), since the thermal broadening effect at the ultra-low temperature $T$ = 60 mK is negligible. The convoluting results are shown as the gray spectra in Fig. 3b-c. The full energy gap is only slightly affected and the thermal broadening effect is minimal even when the temperature increases to 45 K due to the extremely large energy gap at 60 mK. Thus, the experimentally observed rapid filling of the gap does not originate from the thermal broadening, but is most probably caused by the unusual evolution of the superconductivity. It suggests a transformation of the SC gap structure from nodeless to nodal with residual ungapped single-particle density of states. Previous theoretical study reveals that $d$-wave superconducting grains embedded in a metallic matrix can order a global $s$-wave symmetry[43], which is a possible explanation for the temperature evolution of the tunneling spectra. Due to the disorders, the film behaves like $d$-wave superconducting grains with the Josephson coupling, which can be weakened by increasing temperature. At low temperatures, the coupling between different grains is strong enough to order the global nodeless $s$-wave like pairing symmetry, which can be revealed by the U-shaped energy gap with zero residual density of states. At higher temperatures, the coupling of the grains is weakened, and the $d$-wave superconducting gap of an isolate grain can be detected, characterized by the V-shaped gap in the tunneling spectra. We also carried out similar STM/S measurements in another region of the $(La,Pr)_3Ni_2O_7$ thin film (Region 3). This region was treated by pushing the tip towards the sample for around 1 nm and then quickly pulling back. As shown in Extended Data Figs. 3-5, the tunneling spectra show flat-bottom U-shaped energy gap with a similar unconventional temperature evolution.

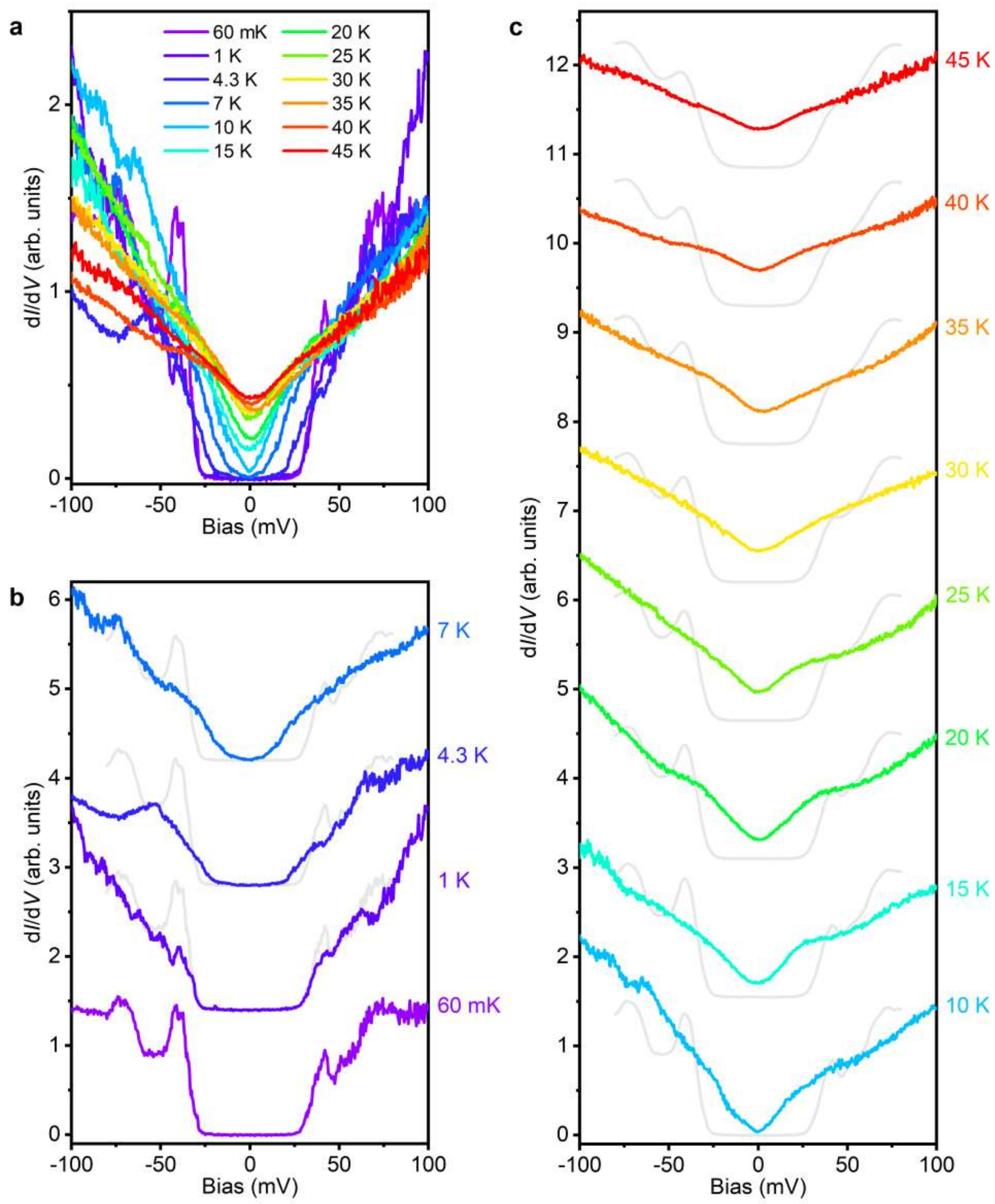

**Fig. 3 | Tunneling spectra measured at different temperatures. a**, The temperature dependence of the tunneling spectra from 60 mK to 45 K measured at the position marked by the red star in Fig.1c ($V_s$ = 100 mV, $I_s$ = 500 pA, $V_{mod}$ = 2 mV). **b-c**, The same spectra in **a** vertically shifted for clarity. The spectra measured from 60 mK to 7 K, from 10 K to 45 K are shown in **b**, **c**, respectively. The gray spectra are obtained by convoluting the spectrum measured at 60 mK by the thermal broadening functions at the corresponding temperatures.

Furthermore, our observations were also repeated in another three-unit-cell thick $(La,Pr)_3Ni_2O_7$ thin film, with a similar La:Pr ratio of about 1.95:1.05, grown on the as-received (001)-oriented $SrLaAlO_4$ substrate by GAE[24,35] (Sample 2). The transport measurements revealed an onset $T_c$ of 48.1 K (Extended Data Fig. 6), and the surface of this film was similarly treated before STM/S measurements by pushing the tip toward the sample for several nanometers and then quickly pulling back. The similar flat-bottom U-shaped energy gap at ultra-low temperatures, the reduction of the energy gap under magnetic field, and the unconventional temperature evolution are shown in Extended Data Figs. 7-8, demonstrating that our observations in the $(La,Pr)_3Ni_2O_7$ thin films are reliable.

**Discussion**

The dependence of the U-shaped energy gap on the magnetic field and temperature is consistent with the behavior of a SC gap, suggesting that the low energy quasiparticle excitations in the $(La,Pr)_3Ni_2O_7$ thin film superconductor are probably dominated by the Bogoliubov quasiparticles with an extremely large nodeless SC gap of around 41.6 meV. Despite the lack of consensus on the mechanism of superconductivity, the theoretical estimate of the SC gap in pressurized R-P bilayer nickelate $La_3Ni_2O_7$ crystals at the γ pocket (derived by the Ni-$3d_{z^2}$ orbital)[5] under certain doping levels can be on the same order as the observed gap in our STM/S experimental results[6]. The consistence of the calculated gap size at the γ pocket[6] and our experimental observation may indicate the dominate role of the γ band for superconductivity in the compressively strained $(La,Pr)_3Ni_2O_7$ thin film at nearly zero pressure, similar to its bulk counterparts in high pressure situation[1].

Here, we would also like to discuss other possible origins for the observed U-shaped energy gap. One possible candidate is the Coulomb blockade, which can induce a similar spectra gap when the charging energy $E_c = e^2/2C$ ($e$ is the electron charge and $C$ is the capacitance) is larger than the thermal energy ($k_BT$)[44,45]. However, our experimental results are inconsistent with this scenario. The U-shaped energy gap is reduced under external magnetic field and rapidly filled with increasing temperatures. At $T$ = 10 K, the energy gap is filled and evolves into V-shaped while the thermal energy ($k_BT \approx 0.86$ meV) is still much smaller than the gap width ($2\Delta$ exceeding 80 meV), indicating a temperature-dependent order parameter rather than a rigid charging energy. Moreover, the energy gap originating from the Coulomb blockade usually exhibits asymmetry about the Fermi level due to the fractional residual charge[46,47], which is inconsistent with the energy-symmetry of the U-shaped gap with coherence peak structures we observed. In addition, the observed U-shaped energy gap is unlikely to be a trivial band gap, which is expected to remain almost unchanged across the temperature range from 60 mK to 10 K. Therefore, the observed U-shaped energy gap is most probably a superconducting gap. To unambiguously confirm the SC origin of the gap, further experimental evidence is still necessary, such as the detection of Josephson tunneling using a SC tip and the direct visualization of vortex lattice under a magnetic field. These experiments demand superior surface quality, which is highly challenging under current conditions due to the inevitable exposure to the atmosphere during the sample transfer process.

Notably, if the observed U-shaped energy gap originates from superconductivity, which is the most possible case, its features would indicate a highly unconventional superconductivity existing in $(La,Pr)_3Ni_2O_7$ thin films grown on $SrLaAlO_4$ substrates. Angle-resolved photoemission spectroscopy (ARPES) studies on a similar system ($(La,Pr,Sm)_3Ni_2O_7$ thin films grown on $SrLaAlO_4$ substrates) reveal a ratio of $\frac{2\Delta}{k_B T_c}$ around 9 (ref. [48]), which is comparable to that of high-$T_c$ cuprate $Bi_2Sr_2CaCu_2O_{8+\delta}$ single crystals[42]. If we assume that the $(La,Pr)_3Ni_2O_7$ thin film in our experiment processes a similar $\frac{2\Delta}{k_B T_c}$ ratio, the observed SC gap $\Delta = 41.6$ meV will correspond to a SC transition temperature $T_c$ around 107 K, which is well above the liquid nitrogen boiling temperature of 77 K. We note that prior to the STM/S measurements, the sample surface was treated by pushing the tip towards the sample for several nanometers and subsequently retracting it rapidly. Besides refreshing the surface, this process may also locally change the out-of-plane lattice constant of the film, which could induce a larger SC gap and a higher "local $T_c$" in the local region of the film than the $T_c$ of the entire film obtained by transport measurements[23,49]. Recently, it was theoretically proposed that introducing an electric field can raise the $T_c$ of the $La_3Ni_2O_7$ thin film to above 80 K at ambient pressure[50]. Furthermore, we found that the tunneling spectra in the tip treated region at low temperatures is robust against the thermal cycling of up to 45 K and then down to the low temperatures (Extended Data Fig. 9), suggesting that the detected extremely large full SC gap and the unconventional temperature dependence of the gap structure in the tunneling spectrum is not related to irreversible thermal processes, such as lattice relaxations.

On the other hand, if we assume that the extremely large SC gap corresponds to the onset $T_c$ obtained by transport measurements in our sample, the $\frac{2\Delta}{k_B T_c}$ ratio would be around 20, which is far from the weak-coupling BCS limit[27] and deeply inside the strong-coupling regime. As a comparison, the $\frac{2\Delta}{k_B T_c}$ ratio is typically about 5-8 for iron-based superconductors[30,51,52], which is much smaller. For most of the cuprate superconductors, the $\frac{2\Delta}{k_B T_c}$ ratio is also much smaller than 20 (ref. [29]). Only very few cuprates, like overdoped $Bi_2Sr_2CuO_{6+\delta}$ with a $T_c$ around 10 K, exhibit $\frac{2\Delta}{k_B T_c}$ ratio larger than 20 (ref. [53]).

In conclusion, we have successfully detected an energy-symmetric, flat-bottom U-shaped energy gap with coherence peak structures and zero residual density of states near the Fermi level in the R-P bilayer nickelate $(La,Pr)_3Ni_2O_7$ thin film grown on the $SrLaAlO_4$ substrate. The unprecedentedly large energy gap raises the potential for $(La,Pr)_3Ni_2O_7$ thin films to achieve a SC transition temperature $T_c$ above the liquid nitrogen boiling temperature at ambient or zero pressure. Our findings offer new insights for understanding the pairing mechanism of the high-$T_c$ superconductivity in nickelate superconductors and other unconventional superconductors.

## Methods

### STM/S measurements

The STM/S experiments were performed on a commercial USM-1600 ultra-high vacuum MBE-STM combined system. All measurements were carried out using a Pt/Ir tip at ultra-low temperatures (base temperature below 70 mK) unless noted otherwise. The STS spectra were acquired by standard lock-in technique at 983 Hz.

### Thermal broadening of the tunneling spectra at high temperatures

In STS measurements, when the local density of states (LDOS) of the STM tip and the tunneling matrix elements remain constant, the tunneling current can be expressed as[54–56]: $I \propto \int_{-\infty}^{+\infty} D(E) \cdot [f(E-eV;T) - f(E;T)] \cdot \mathrm{d}E$, where $D(E)$ is the LDOS of the sample at the energy $E$, $V$ is the bias voltage, $f(E;T) = \frac{1}{1+\exp(\frac{E-E_\mathrm{F}}{k_\mathrm{B}T})}$ is the Fermi-Dirac distribution function, and $E_\mathrm{F}$ is the Fermi level. Thus the differential conductance ($\frac{\mathrm{d}I}{\mathrm{d}V}$) can be obtained by calculating the derivative of the tunneling current $I$ with respect to the bias voltage $V$: $\frac{\mathrm{d}I}{\mathrm{d}V} \propto \int_{-\infty}^{+\infty} D(E) \cdot \frac{\mathrm{d}}{\mathrm{d}V} f(E-eV;T) \cdot \mathrm{d}E = e\int_{-\infty}^{+\infty} D(E) \cdot \frac{1}{4k_\mathrm{B}T}\cosh^{-2}\left(\frac{E-E_\mathrm{F}-eV}{2k_\mathrm{B}T}\right) \cdot \mathrm{d}E$. So $\frac{\mathrm{d}I}{\mathrm{d}V} = C\int_{-\infty}^{+\infty} D(E) \cdot F(E;V,T) \cdot \mathrm{d}E$, where $C$ is a proportionality coefficient and $F(E;V,T) = \frac{1}{4k_\mathrm{B}T}\cosh^{-2}\left(\frac{E-E_\mathrm{F}-eV}{2k_\mathrm{B}T}\right)$ is the normalized thermal broadening function ($\int_{-\infty}^{+\infty} F(E;V,T) \cdot \mathrm{d}E = 1$). At ultra-low temperatures ($T \to 0$), $F(E;V,T) \to \delta(E-E_\mathrm{F}-eV)$, which means that $\frac{\mathrm{d}I}{\mathrm{d}V} \to C \cdot D(E_\mathrm{F}+eV)$ is proportional to the LDOS of the sample at $E = E_\mathrm{F} + eV$. However, as the temperature increases, peak broadening of the function $F(E;V,T)$ also increases accordingly, which means that the convolution of LDOS by the thermal broadening function should be taken into consideration at high temperatures. In practice, the thermal broadened spectra at high temperatures (gray curves in Fig. 3b-c, Extended Data Fig. 5c and Extended Data Fig. 7c) are obtained by numerically convoluting the spectrum measured at lowest temperature (60 mK, 55 mK, 60 mK in Fig. 3b-c, Extended Data Fig. 5c, Extended Data Fig. 7c, respectively) by the thermal broadening function. Similar method was also used in previous STS studies[57–61].

### Charge transfer gap revealed by tunneling spectra

We measured the tunneling spectra in a wider energy range on the surface of the $(La,Pr)_3Ni_2O_7$ film after the tip treatment (Extended Data Fig. 10). The tunneling spectra show a huge gap feature beyond the U-shaped energy gap around the Fermi level, which may correspond to the charge transfer gap and indicates the presence of Mottness[17–19,36]. The charge transfer energy ($\Delta_{CT}$, determined by measuring the difference between the onset energies of the upper Hubbard band and the charge transfer band) is approximately 1.6 - 1.7 eV, similar to the previous results[17,20,36].

## Data Availability

All data that support the findings of this study are available from the corresponding author on reasonable request.

## Acknowledgements

We acknowledge Jinfeng Jia, Zhuoyu Chen and Qi-Kun Xue for offering us the high-$T_c$ nickelate $(La,Pr)_3Ni_2O_7$ thin films. We thank Tao Xiang for insightful discussions. This work was financially supported by the National Natural Science Foundation of China (Grant No. 12488201), the Quantum Science and Technology-National Science and Technology Major Project (2021ZD0302403), the National Key Research and Development Program of China (Grant No. 2025YFA1411300). Z.W. is supported by the U.S. Department of Energy, Basic Energy Sciences (Grant No. DE-FG02-99ER45747).

## Author contributions

J.W. conceived and supervised the project. Z.L., T.W., W.R. and Y.L. carried out the STM/S experiments. H.J. and Z.X. performed the transport measurements. Z.L., T.W., Z.W. and J.W. analyzed the experimental data. Z.L, T.W., Z.W. and J.W. wrote the manuscript.

## Competing interests

The authors declare no competing interests.

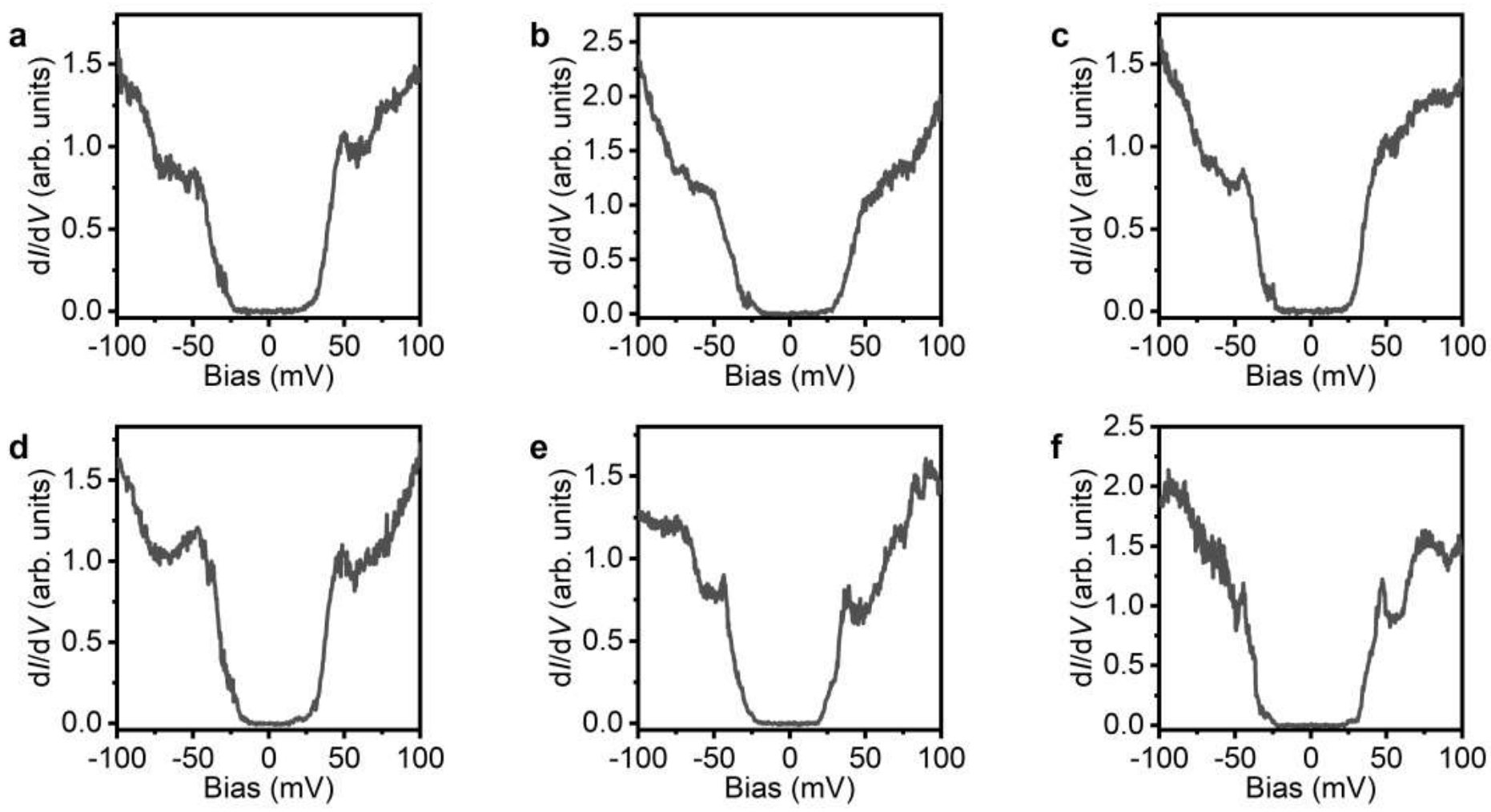

**Extended Data Fig. 1 | More tunneling spectra measured in Region 1. a-f**, Tunneling spectra measured at different positions in Region 1 ($V_s$ = 100 mV, $I_s$ = 500 pA, $V_{mod}$ = 2 mV). Measurement temperatures are $T$ = 50 mK (**a**-**c**) and $T$ = 60 mK (**d**-**f**), respectively.

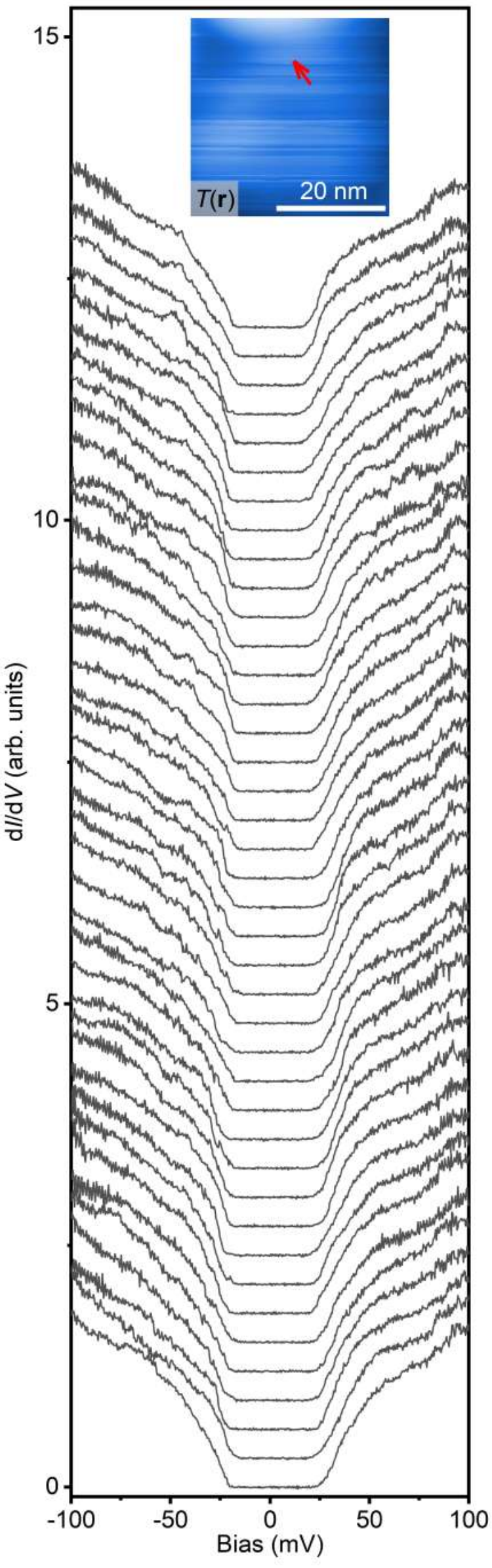

**Extended Data Fig. 2 | Flat-bottom U-shaped energy gap measured in another region of the $(La,Pr)_3Ni_2O_7$ thin film (Sample 1).** The tunneling spectra measured along the red arrow shown in the inset without magnetic field ($V_s$ = 100 mV, $I_s$ = 500 pA, $V_{mod}$ = 2 mV, $T$ = 50 mK). The spectra are vertically shifted for clarity. The inset shows a topographic image of the $(La,Pr)_3Ni_2O_7$ thin film in Region 2 (36.57 × 36.57 nm$^2$, $V_s$ = 8 V, $I_s$ = 100 pA) after the tip treatment.

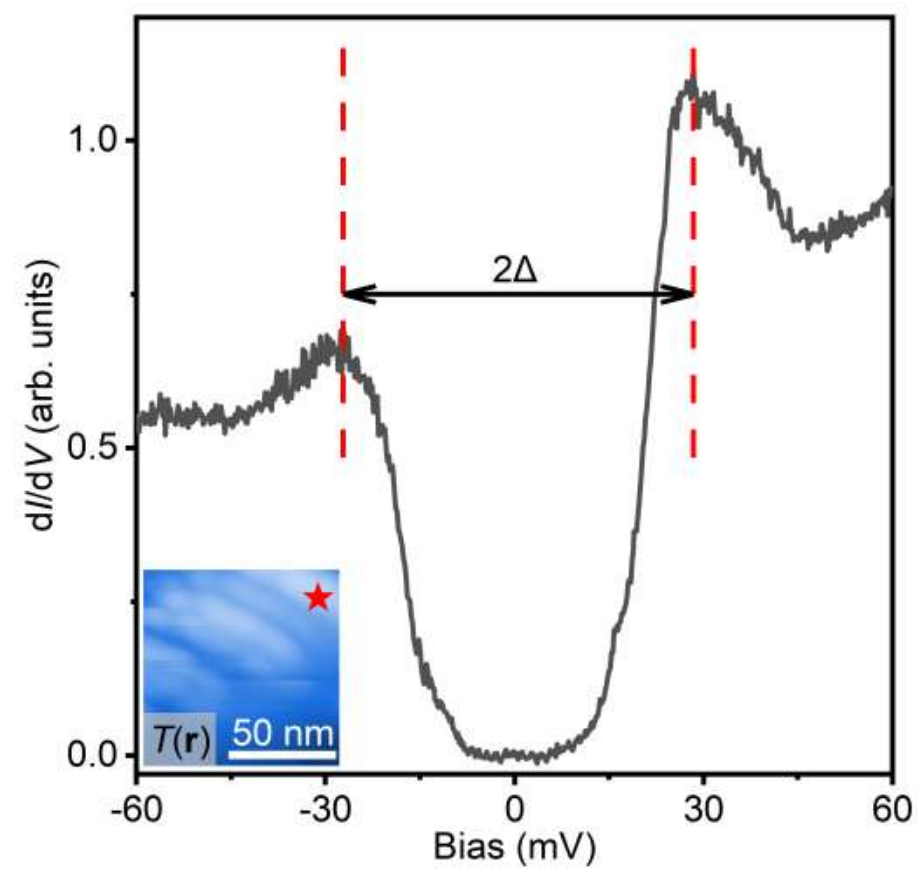

**Extended Data Fig. 3 | Tunneling spectrum with flat-bottom U-shaped energy gap structure measured in another region of the (La,Pr)$_3$Ni$_2$O$_7$ thin film (Sample 1).** A typical tunneling spectrum measured on the position marked by the red star in the inset ($V_s$ = 60 mV, $I_s$ = 500 pA, $V_{mod}$ = 1 mV, $T$ = 50 mK). The gap size (Δ) can be determined to be around 27.8 meV by half of the separation in energy between the two coherence peaks. The inset shows a topographic image of the (La,Pr)$_3$Ni$_2$O$_7$ thin film in Region 3 (91.43 × 91.43 nm$^2$, $V_s$ = 8 V, $I_s$ = 10 pA) after the tip treatment.

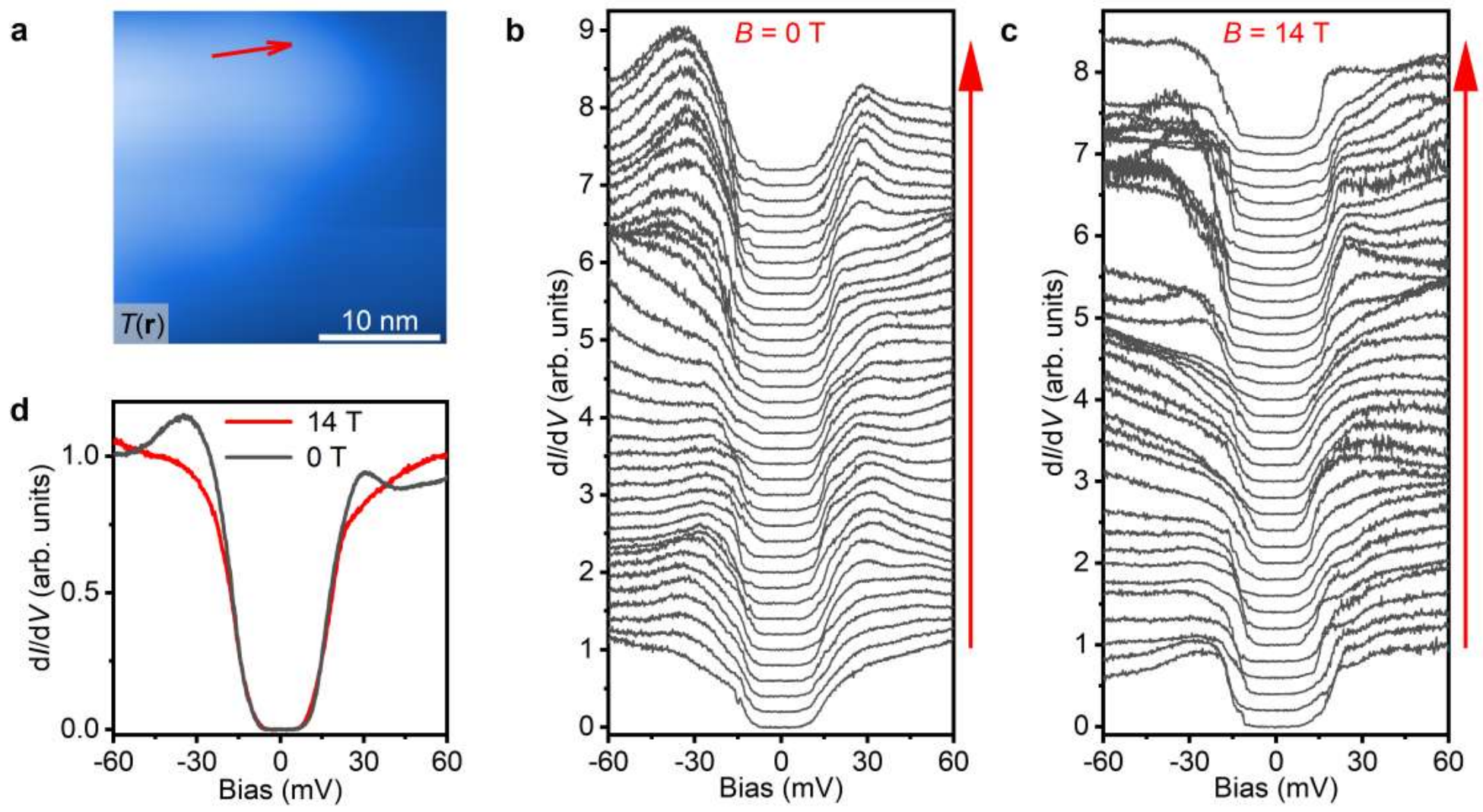

**Extended Data Fig. 4 | Suppression of the coherence peaks under the magnetic field *B* = 14 T perpendicular to the surface.** **a**, A topographic image of the $(La,Pr)_3Ni_2O_7$ thin film in Region 3 ($27.43 \times 27.43$ nm$^2$, $V_s$ = 8 V, $I_s$ = 100 pA). **b**, The tunneling spectra measured along the red arrow shown in **a** without magnetic field ($V_s$ = 60 mV, $I_s$ = 500 pA, $V_{mod}$ = 1 mV, $T$ = 50 mK). The spectra are vertically shifted for clarity. **c**, The tunneling spectra measured along the same arrow in **a** under the magnetic field $B$ = 14 T perpendicular to the surface ($V_s$ = 60 mV, $I_s$ = 500 pA, $V_{mod}$ = 1 mV, $T$ = 50 mK). The spectra are vertically shifted for clarity. **d**, The spatially averaged spectra of the spectra shown in **b** ($B$ = 0 T, black curve) and **c** ($B$ = 14 T, red curve). The coherence peaks are obviously suppressed by the magnetic field.

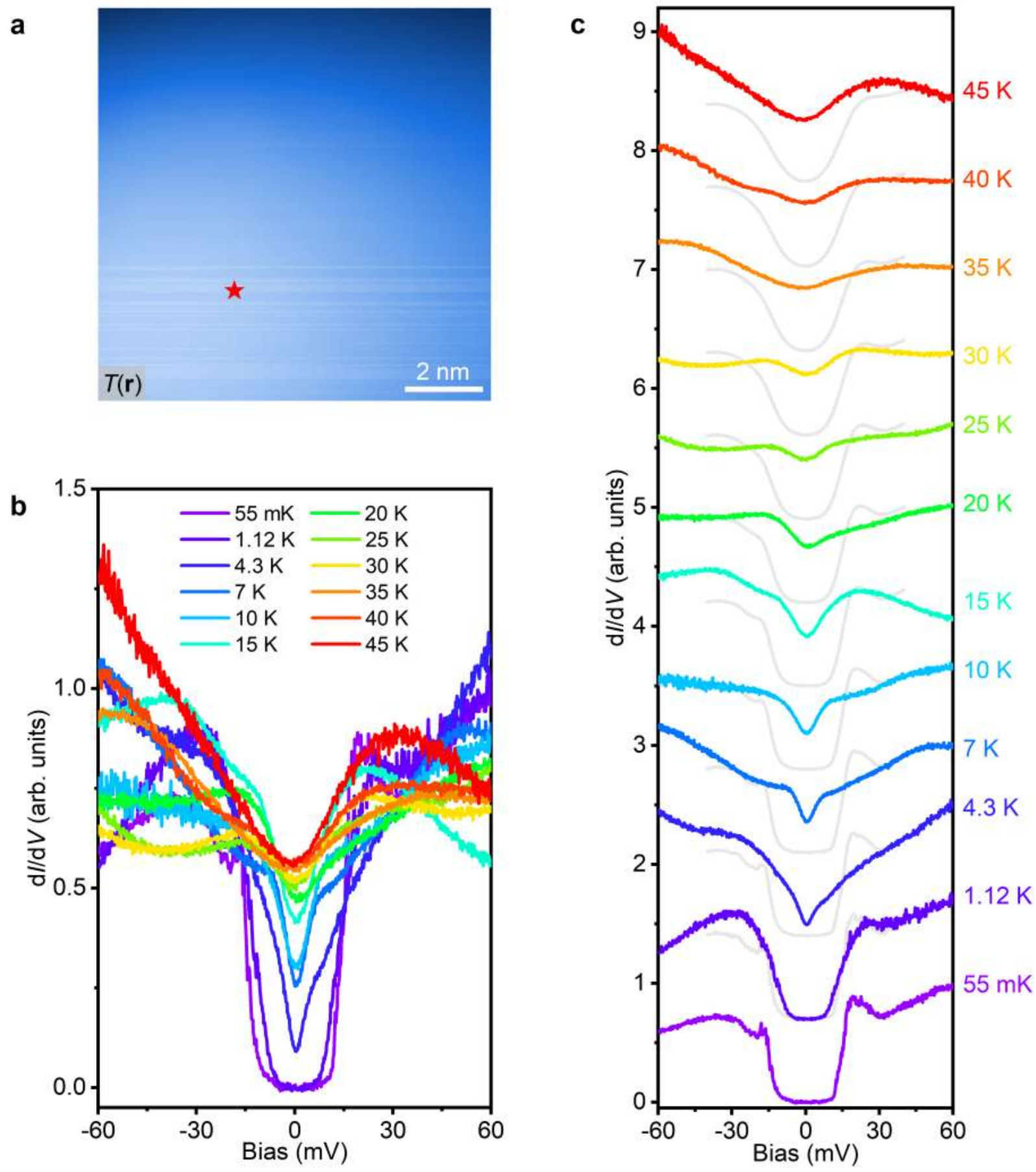

**Extended Data Fig. 5 | Tunneling spectra measured at different temperatures in Region 3. a**, A topographic image of the $(La,Pr)_3Ni_2O_7$ thin film in Region 3 (10.06 × 10.06 nm$^2$, $V_s$ = 8 V, $I_s$ = 100 pA). **b**, The temperature dependence of the tunneling spectra from 55 mK to 45 K measured on the position marked by the red star in **a** ($V_s$ = 60 mV, $I_s$ = 500 pA, $V_{mod}$ = 1 mV). The gap size (Δ) is around 18.6 meV at 55 mK. **c**, The same spectra in **b** vertically shifted for clarity. The gray spectra are obtained by convoluting the spectrum measured at 55 mK by the thermal broadening functions at higher temperatures for comparison.

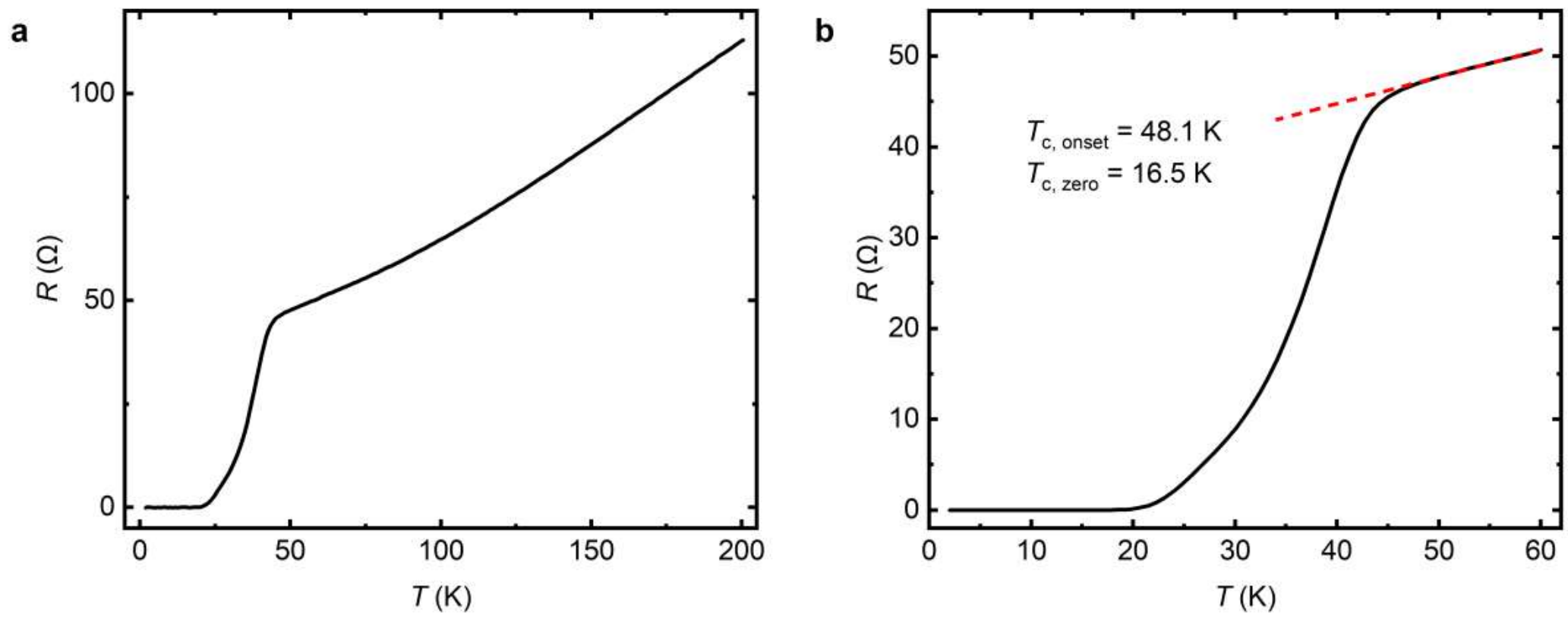

**Extended Data Fig. 6 | The *R*-*T* curve of another three-unit-cell thick $(La,Pr)_3Ni_2O_7$ thin film (Sample 2) grown on the as-received (001)-oriented $SrLaAlO_4$ substrate by GAE measured before STM/S experiments. b** is the zoom-in *R*-*T* curve of **a**, showing the onset SC $T_c$ of 48.1 K and the zero resistance $T_c$ of 16.5 K. The La:Pr ratio of this nickelate thin film is about 1.95:1.05.

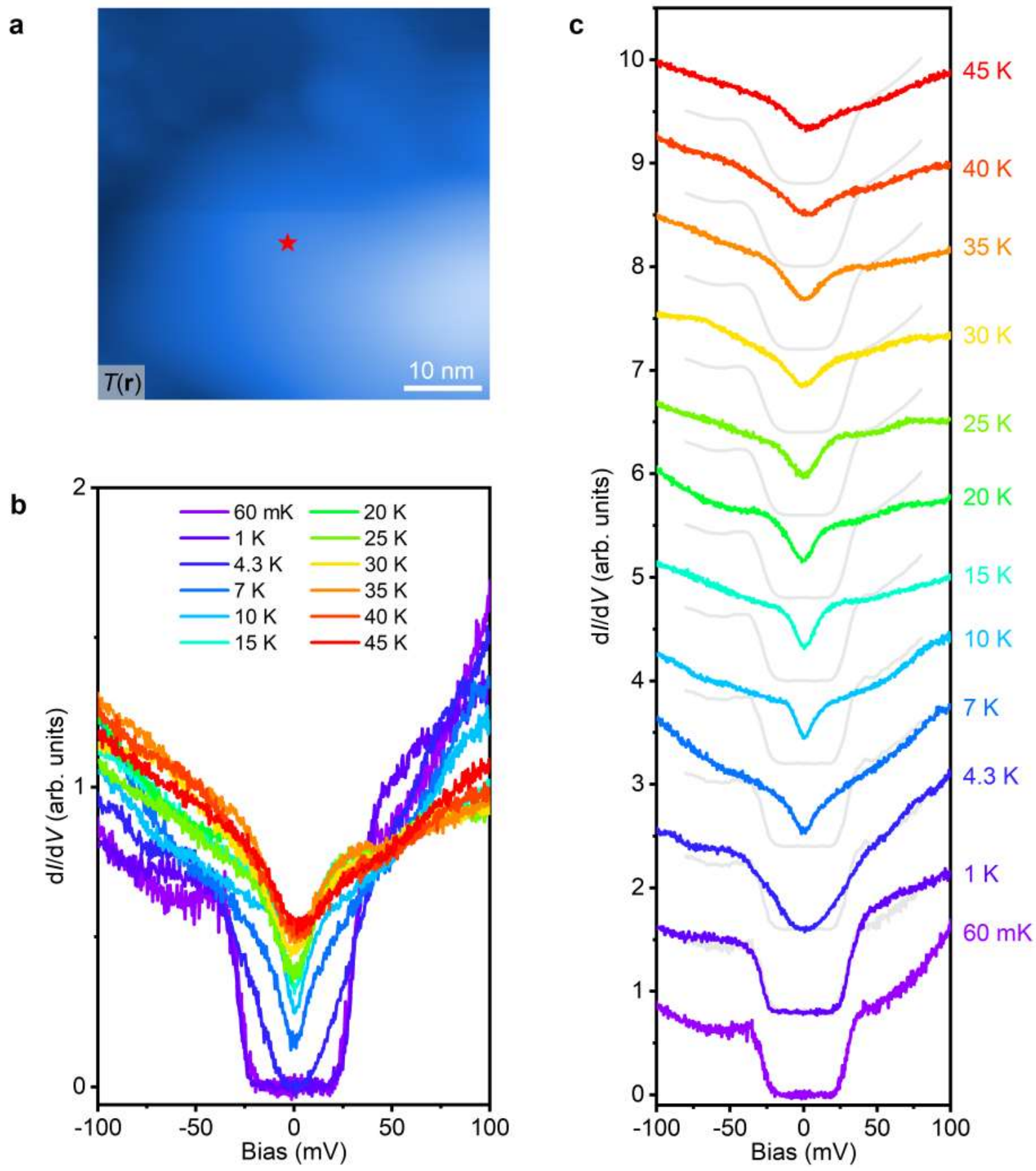

**Extended Data Fig. 7 | Tunneling spectra measured at different temperatures in a region obtained by tip treatment on Sample 2. a**, A topographic image of the $(La,Pr)_3Ni_2O_7$ thin film in Region 4 ($50.29 \times 50.29$ nm$^2$, $V_s = 8$ V, $I_s = 100$ pA) after the tip treatment. **b**, The temperature dependence of the tunneling spectra from 60 mK to 45 K measured on the position marked by the red star in **a** ($V_s = 100$ mV, $I_s = 500$ pA, $V_{mod} = 2$ mV). The gap size (Δ) is around 35.2 meV at 60 mK. **c**, The same spectra in **b** vertically shifted for clarity. The gray spectra are obtained by convoluting the spectrum measured at 60 mK by the thermal broadening functions at higher temperatures for comparison.

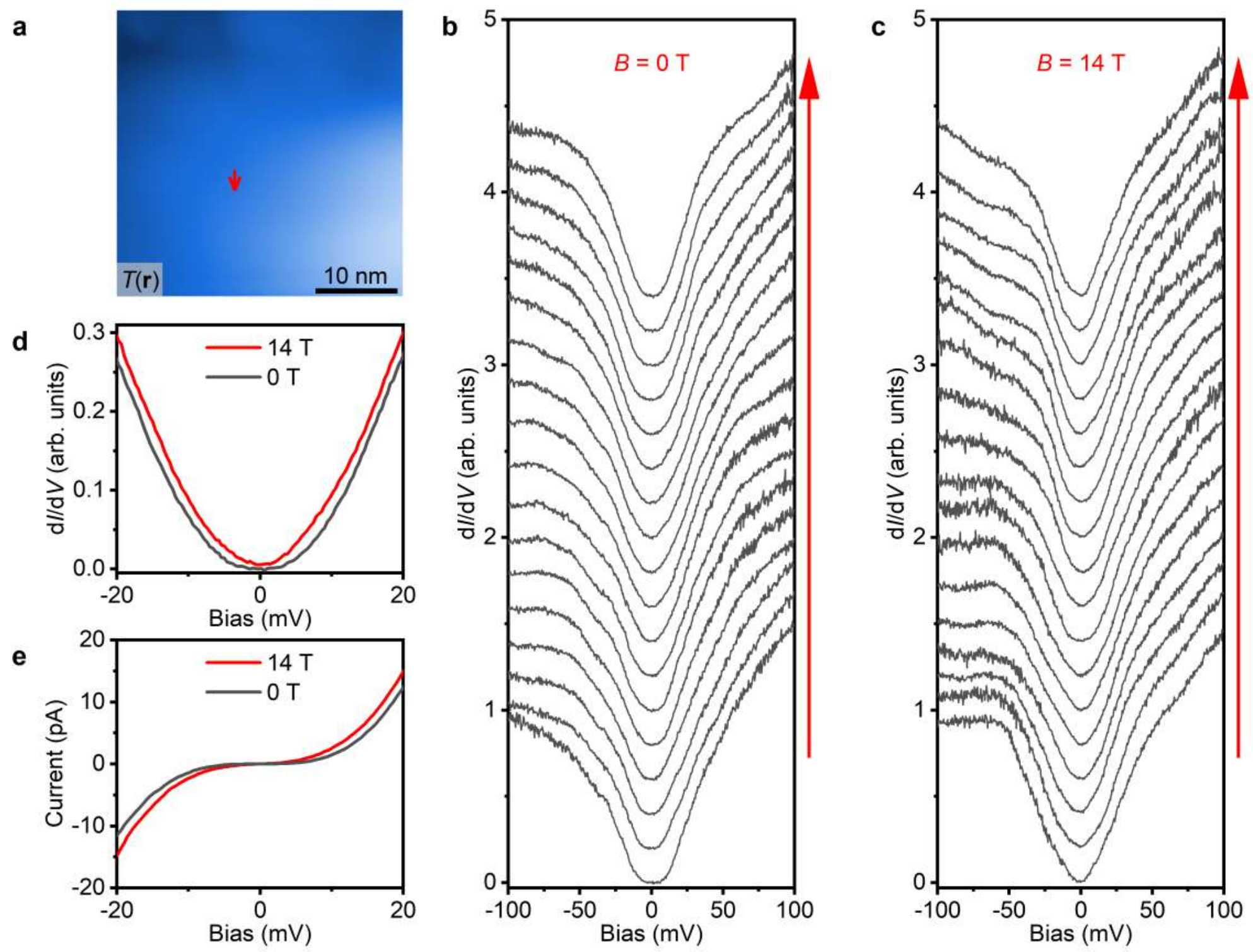

**Extended Data Fig. 8 | Reduction of the energy gap under the magnetic field $B$ = 14 T perpendicular to the surface (Sample 2).** **a**, A topographic image of the $(La,Pr)_3Ni_2O_7$ thin film in Region 4 (35 × 35 $nm^2$, $V_s$ = 8 V, $I_s$ = 100 pA) after the tip treatment. **b**, The tunneling spectra measured along the red arrow shown in **a** without magnetic field ($V_s$ = 100 mV, $I_s$ = 500 pA, $V_{mod}$ = 2 mV, $T$ = 4.3 K). The spectra are vertically shifted for clarity. **c**, The tunneling spectra measured along the same arrow in **a** under the magnetic field $B$ = 14 T perpendicular to the surface ($V_s$ = 100 mV, $I_s$ = 500 pA, $V_{mod}$ = 2 mV, $T$ = 4.3 K). The spectra are vertically shifted for clarity. **d**, The zoom-in spatially averaged spectra of the spectra shown in **b** ($B$ = 0 T, black curve) and **c** ($B$ = 14 T, red curve). The red curve under the magnetic field is more like V-shaped than the black curve under zero field. **e**, The zoom-in spatially averaged $I$-$V$ curves of the $I$-$V$ curves simultaneously measured with the spectra shown in **b** ($B$ = 0 T, black curve) and **c** ($B$ = 14 T, red curve). The red tunneling $I$-$V$ curve under the magnetic field $B$ = 14 T is less flat near the Fermi level than the black $I$-$V$ curve under zero field, which probably indicates the reduction of the superconductivity under the magnetic field.

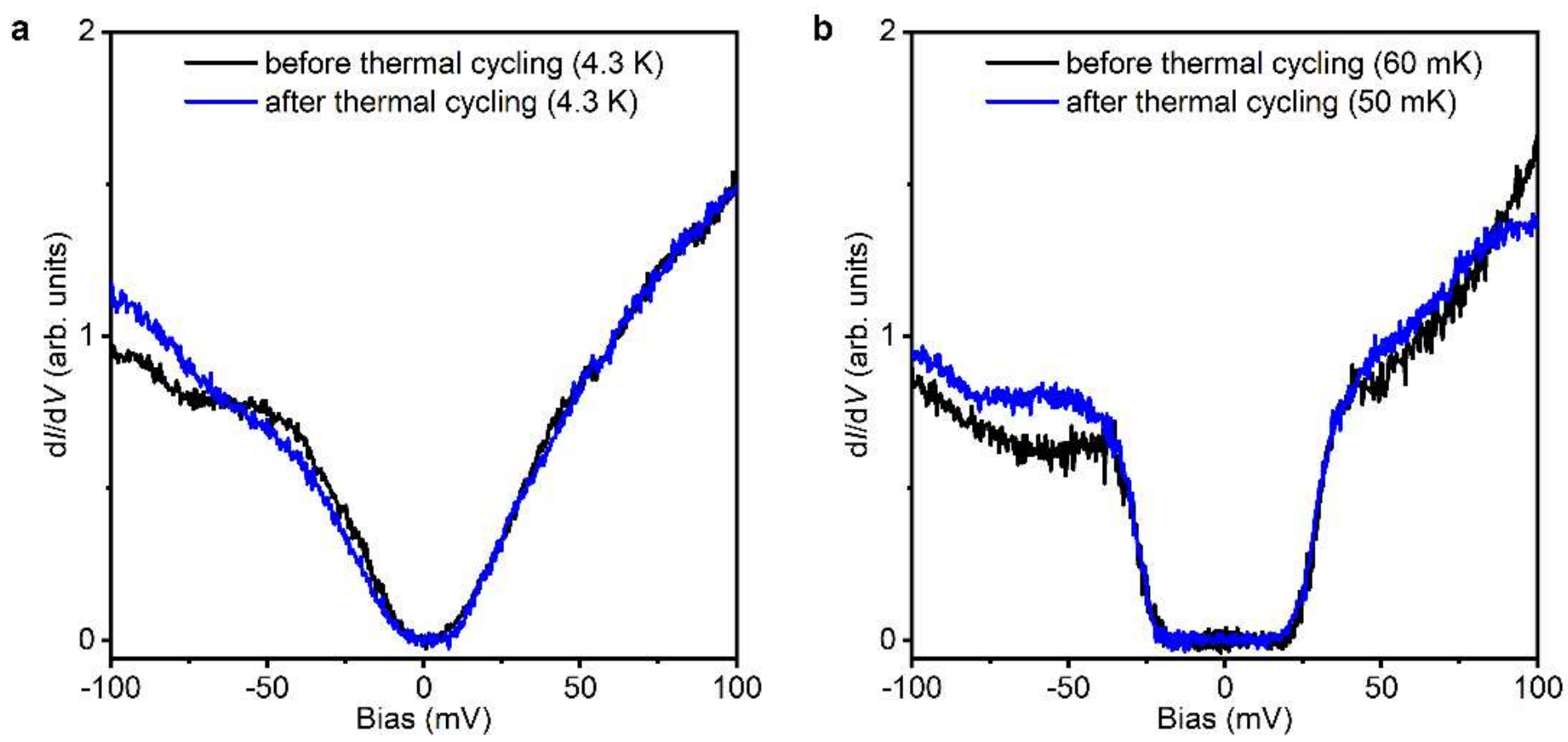

**Extended Data Fig. 9 | Robust tunneling spectra against the thermal cycling (Sample 2). a**, Tunneling spectra measured at 4.3 K before (black) and after (blue) the thermal cycling (the measurement interval exceeds 25 hours). **b**, Tunneling spectra measured at ultra-low temperatures before (black, 60 mK) and after (blue, 50 mK) the thermal cycling (the measurement interval exceeds 115 hours). The tunneling spectra were measured at the same position shown as the red star in Extended Data Fig. 7. Measurement conditions: $V_s$ = 100 mV, $I_s$ = 500 pA, $V_{mod}$ = 2 mV.

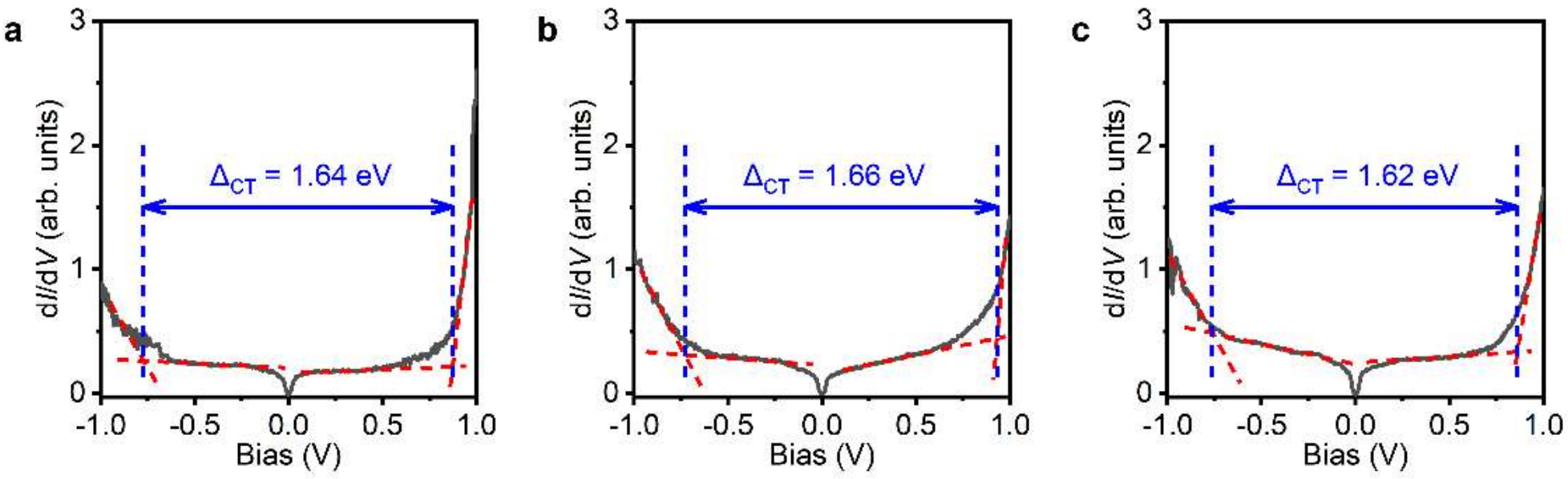

**Extended Data Fig. 10 | Charge transfer energy revealed by the tunneling spectra (Sample 2). a-c**, Tunneling spectra measured at different positions in another region (Region 5) of Sample 2 after the tip treatment ($V_s$ = 1 V, $I_s$ = 500 pA, $V_{mod}$ = 10 mV, $T$ = 50 mK). The charge transfer energy ($\Delta_{CT}$) is approximately 1.6 – 1.7 eV. Due to the large modulation voltage ($V_{mod}$) used in lock-in measurements, the energy resolution (~ 2.5 $eV_{mod}$ = 25 meV) is not enough to precisely resolve the U-shaped energy gap structure around the Fermi level.

# Supplementary Information for

# Observation of flat-bottom U-shaped energy gap in high-$T_c$ nickelate $(La,Pr)_3Ni_2O_7$ thin films

Zhen Liang[1,2,3,#], Tianheng Wei[1,2,#], Wei Ren[1,2,#], Haoran Ji[1,2], Zheyuan Xie[1,2], Yanzhao Liu[4], Ziqiang Wang[5] & Jian Wang[1,2,3,6,*]

[1]International Center for Quantum Materials, School of Physics, Peking University, Beijing 100871, China

[2]Beijing Key Laboratory of Quantum Devices, Peking University, Beijing 100871, China

[3]Hefei National Laboratory, Hefei 230088, China

[4]Quantum Science Center of Guangdong-Hong Kong-Macao Greater Bay Area, Shenzhen, 518045, China

[5]Department of Physics, Boston College, Chestnut Hill, MA 0246, USA

[6]Collaborative Innovation Center of Quantum Matter, Beijing 100871, China

[#]These authors contributed equally.

[*]Corresponding to: jianwangphysics@pku.edu.cn (J.W.).

**The rough surface of the $(La,Pr)_3Ni_2O_7$ thin film grown on the $SrLaAlO_4(001)$ substrates before the STM tip treatment**

Before the STM tip treatment, the original surface of the $(La,Pr)_3Ni_2O_7$ thin film show the roughness at nanometer scale (Fig. S1), possibly due to the surface degradation in atmosphere during the sample transfer process.

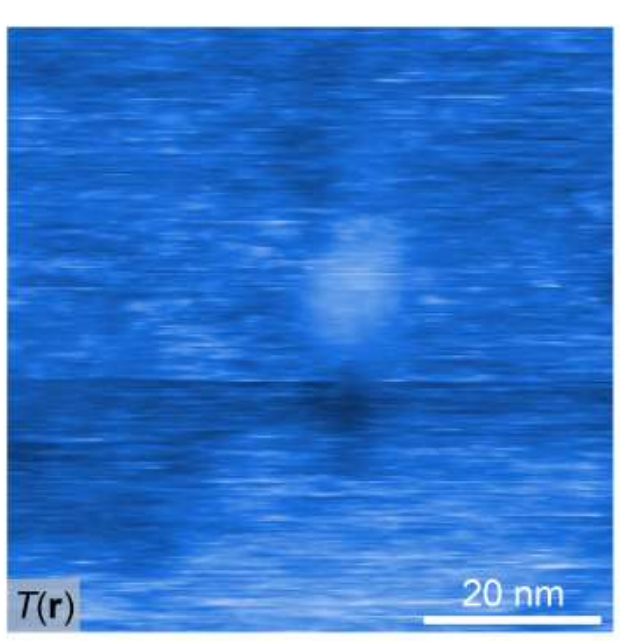

**Fig. S1 | Original surface of the $(La,Pr)_3Ni_2O_7$ thin film before the tip treatment.** A typical topographic image of the $(La,Pr)_3Ni_2O_7$ thin film before surface treatment ($68.57 \times 68.57$ nm$^2$, $V_s = 8$ V, $I_s = 500$ pA).

**Relationship between the gap size and the U-to-V transition temperature of the tunneling spectra**

From the temperature dependent measurements (Fig. 3, Extended Data Fig. 5, and Extended Data Fig. 7), we extracted the values of the gap sizes at ultra-low temperatures, and the transition temperatures at which the tunneling spectra evolve from U-shaped to V-shaped. The results are summarized in Fig. S2, revealing that a larger gap size corresponds to a higher U-to-V transition temperature of the tunneling spectra.

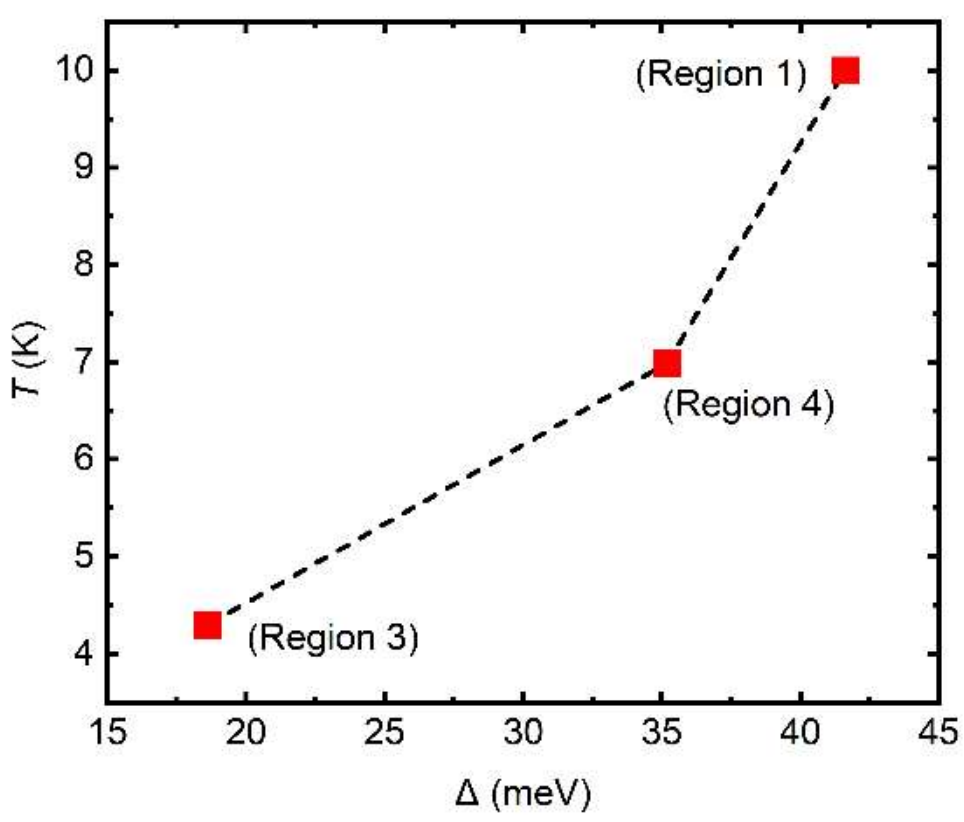

**Fig. S2 | Correlation between the gap size and the U-to-V transition temperature of the tunneling spectra.** The results are extracted from Fig. 3, Extended Data Fig. 5,

and Extended Data Fig. 7.

**Fitting for the gap edge of the ultra-low-temperature tunneling spectrum**

At the ultra-low temperature (60 mK), the tunneling spectrum (Fig. 1c) near the Fermi level can be well fitted by an exponential function, as shown in Fig. S3, further confirming the nodeless U-shaped gap characteristic.

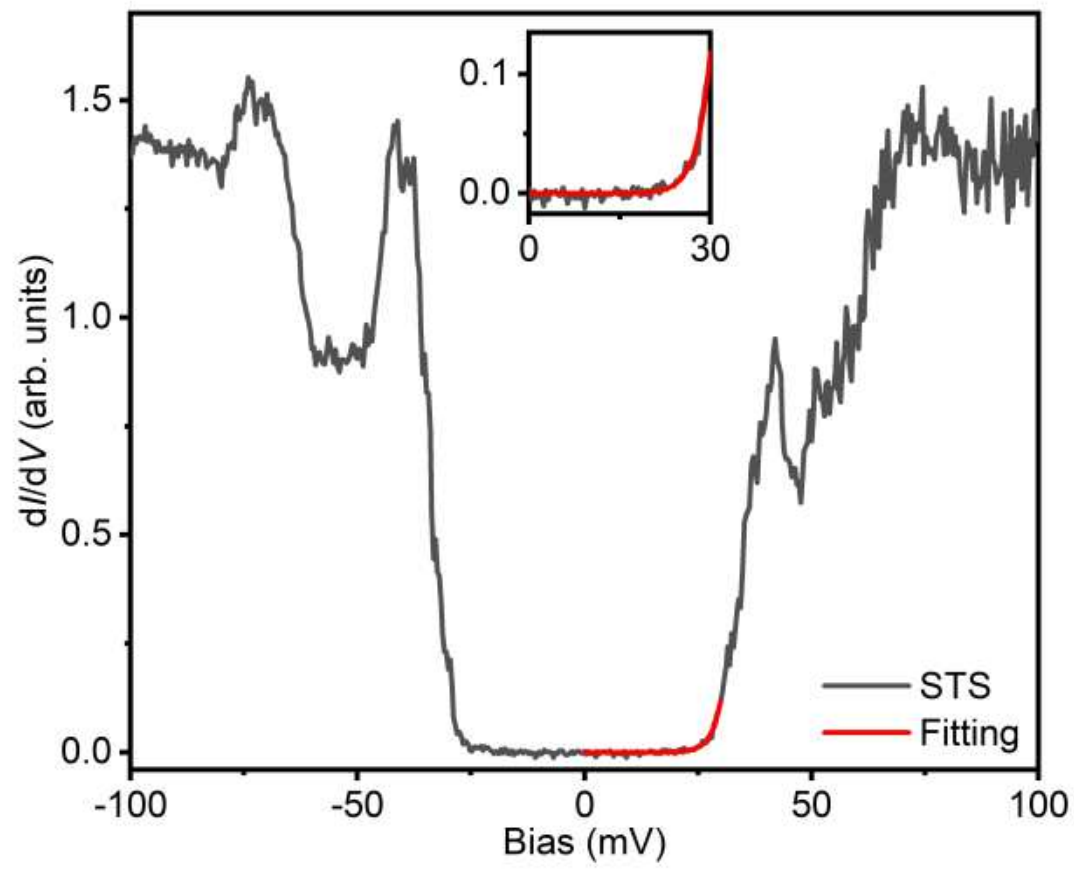

**Fig. S3 | Exponential fitting for the gap edge of the tunneling spectrum measured at 60 mK.** The black curve is the tunneling spectrum measured at 60 mK shown in Fig. 1c. The red curve shows the theoretical fitting result for the black curve within 0-30 mV, using an exponential function $y = A*\exp(x/B)$. Fitting parameters: $A = 1.0822*10^{-7}$ (arb. units); $B = 2.1575$ mV. The inset shows the zoom-in tunneling spectrum and the fitting curve.